\documentclass[a4paper,aps,prd,10pt,preprintnumbers,showpacs,twocolumn,superscriptaddress,nofootinbib,amsmath,amssymb]{revtex4-1}

\usepackage{orcidlink,graphicx,cmap}
\usepackage{cmap}
\usepackage[utf8]{inputenc}
\usepackage[T1]{fontenc}

\def\re#1{Re(#1)}
\def\im#1{Im(#1)}

\begin{document}
\title{Gravitational Spectra and Wave Propagation in Regular Black Holes Supported by a Dehnen Halo}
\author{Bekir Can Lütfüoğlu}
\email{bekir.lutfuoglu@uhk.cz}
\affiliation{Department of Physics, Faculty of Science, University of Hradec Králové, Rokitanského 62/26, 500 03 Hradec Králové, Czech Republic}

\author{Abubakir~Shermatov}
\email{shermatov.abubakir98@gmail.com}
\affiliation{Institute of Fundamental and Applied Research, National Research University TIIAME, Kori Niyoziy 39, Tashkent 100000, Uzbekistan}
\affiliation{University of Tashkent for Applied Sciences, Str. Gavhar 1, Tashkent 100149, Uzbekistan}
\affiliation{Tashkent State Technical University, Tashkent 100095, Uzbekistan}

\author{Javlon~Rayimbaev}
\email{javlon@astrin.uz}
\affiliation{National University of Uzbekistan, Tashkent 100174, Uzbekistan}

\author{Muhammad~Matyoqubov} \email{m_matyoqubov@mamunedu.uz}   \affiliation{Mamun University Bolkhovuz street 2, Khiva 220900, Uzbekistan} 

\author{Otaboyev~Sirajiddin}
\email{sirajiddin@urdu.uz}
\affiliation{Urgench State University, Kh. Alimjan street 14, Urgench 221100, Uzbekistan}


\begin{abstract}
We investigate gravitational perturbations, quasinormal modes, grey-body factors, and absorption cross-sections of the recently proposed regular and asymptotically flat black hole supported by a Dehnen-type dark-matter halo. This geometry provides a remarkably simple analytic model of supermassive black holes embedded in galactic environments, having a lapse function
$
f(r)=1-2 M r^{2}/(r+a)^{3},
$
[R. A. Konoplya, A. Zhidenko,  2511.03066]. The regularizing parameter $a$ is the characteristic scale of the halo. We compute the quasinormal spectrum for both axial ``up'' and ``down'' perturbations using the WKB method and verify the results through time-domain integration. The two sectors are no longer isospectral, and the deviations grow with the halo scale parameter. The grey-body factors and absorption cross-sections are extracted via standard scattering boundary conditions and the WKB approach, and their behaviour is fully consistent with the structure of the effective potentials. Altogether, our analysis demonstrates that a dark-matter halo imprint induces modifications in the gravitational response, while the employed approximation schemes remain sufficiently accurate for quantitative predictions. At asymptotically late times, the presence of the halo does not alter the Price–law decay, which remains identical to that of a Schwarzschild black hole in vacuum.
\end{abstract}

\maketitle

\section{Introduction}

The calculation of quasinormal modes (QNMs) \cite{Kokkotas:1999bd, Konoplya:2011qq, Berti:2009kk, Bolokhov:2025uxz}  and grey-body factors (GBFs) \cite{Page:1976df, Kanti:2004nr} of black holes plays a central role in contemporary gravitational physics. QNMs describe the characteristic oscillations governing the relaxation of perturbed black holes, providing a direct theoretical link to the ringdown signals observed in gravitational-wave experiments \cite{LIGOScientific:2016aoc, LIGOScientific:2017vwq, LIGOScientific:2020zkf}. Their spectrum encodes detailed information about the underlying geometry and field content, making QNMs one of the most sensitive probes of general relativity and its possible extensions in the strong-gravity regime. Complementarily, GBFs determine the frequency-dependent transmission of radiation through the curved spacetime potential barrier surrounding a black hole, shaping both the spectrum of Hawking radiation and the propagation of classical waves \cite{Hawking:1975vcx, Page:1976df}. Together, QNMs and GBFs offer a comprehensive description of black-hole response to perturbations, with wide-ranging implications for stability analyses, tests of the no-hair conjecture, precision gravitational-wave astronomy, and models of quantum or exotic compact objects. As observational sensitivities increase, accurate theoretical predictions for these quantities become indispensable for interpreting current and future data.

Building on this growing awareness that realistic black holes are embedded in complex astrophysical settings rather than isolated in a vacuum, an extensive body of work has emerged over the past few years examining how various forms of surrounding matter modify black-hole perturbations, scattering potentials, and quasinormal ringing. In particular, models of dark-matter halos—both particle-like and effective fluid descriptions—have motivated detailed analyses of waveform distortions, shifted quasinormal spectra, long-lived modes, potential instabilities, and modified GBFs. These studies reveal that even relatively weak environmental densities can leave characteristic imprints on the propagation of fields and on the observable ringdown signal, while denser or more structured halos may generate qualitatively new dynamical features. The resulting literature on black-hole perturbations in non-vacuum spacetimes is now vast and rapidly expanding, especially in the context of dark-matter environments \cite{Konoplya:2021ube,Dubinsky:2025fwv,Feng:2025iao,Mollicone:2024lxy,Tovar:2025apz,Lutfuoglu:2025kqp,Pezzella:2024tkf,Chakraborty:2024gcr,Liu:2024bfj,Liu:2024xcd,Zhao:2023tyo,Malik:2025czt,Daghigh:2022pcr,Zhang:2021bdr,Pathrikar:2025sin,Hamil:2025pte,Rincon:2025buq}, reflecting the increasing importance of incorporating realistic astrophysical surroundings into precision tests of gravity.

Recently, a particularly intriguing development in this direction has been the construction of a new class of regular, asymptotically flat black-hole geometries obtained by coupling Einstein gravity to an anisotropic fluid that effectively models a galactic dark-matter halo~\cite{Konoplya:2025ect}. Unlike many earlier attempts—most of which rely on numerical integration, special ansätze, or matter sources that violate standard energy conditions—this framework yields closed-form analytic solutions for a broad class of matter profiles, including the widely used Dehnen-type distributions. Remarkably, the resulting metrics are free of curvature singularities and exhibit a smooth transition from an exterior region that closely resembles Schwarzschild to an interior domain characterized by an effective de Sitter core. This behaviour is achieved without introducing exotic matter, while maintaining the weak energy condition, thereby placing these models on a firm physical footing. As such, these solutions offer a realistic and self-consistent description of supermassive black holes residing in galactic environments and provide a valuable laboratory for exploring how astrophysical matter distributions influence spacetime geometry, quasinormal spectra, and associated observational signatures.

Here, we are interested in the analysis of quasinormal spectra of the above regular black hole supported by the galactic halo. While quasinormal frequencies of test fields in the Konoplya-Zhidenko regular black hole \cite{Konoplya:2025ect} were recently investigated in \cite{Bolokhov:2025fto}, the physically most relevant sector—gravitational perturbations—has remained unexplored. In this work, we fill this gap by computing the gravitational quasinormal spectrum using high-order WKB methods and validating the results through time-domain evolution. In addition, we determine the corresponding GBFs and absorption cross-sections, thereby providing the comprehensive characterization of gravitational perturbations in this regular and asymptotically flat  Dehnen-supported black-hole spacetime.

The paper is organized as follows. In Sec.~\ref{sec:wavelike} we introduce the regular Dehnen–supported black-hole spacetime, summarize its main geometric properties, and present the two axial gravitational master equations together with their effective potentials. In Sec.~\ref{sec:methodgsforQNMcal} we describe the computational framework used to obtain the quasinormal spectra, combining the higher-order WKB method with Padé approximants and an independent time-domain evolution scheme. The numerical results for the quasinormal frequencies of both ``up'' and ``down'' perturbations, including their dependence on the halo scale parameter and multipole number, are analysed and interpreted in Sec.~\ref{sec:QNMs}. In Sec.~\ref{sec:cross-sections} we turn to the scattering problem, computing the grey–body factors and absorption cross–sections and discussing their relation to the underlying potentials and to the quasinormal spectrum. Finally, Sec.~\ref{sec:Conclusions} contains our conclusions and an outlook for future developments.

\section{Black hole metric and its gravitational perturbations}\label{sec:wavelike}

The spacetime investigated in this work belongs to the family of regular, asymptotically flat black-hole geometries supported by a Dehnen-type dark-matter profile, first constructed in~\cite{Konoplya:2025ect}. The solution arises from Einstein gravity coupled to an anisotropic fluid that effectively models a galactic halo, and—remarkably—admits a closed-form expression for the metric. The line element can be written in the usual static and spherically symmetric form,
\begin{equation}\label{metric}
ds^{2}=-f(r)\,dt^{2}+\frac{dr^{2}}{f(r)}+r^{2}\left(d\theta^{2}+\sin^{2}\theta\,d\phi^{2}\right),
\end{equation}
with the metric function
\begin{equation}\label{fmetric}
f(r)=1-\frac{2M r^{2}}{(r+a)^{3}}\,.
\end{equation}
Here, $M$ represents the ADM mass, while the parameter $a>0$ sets the characteristic scale of the surrounding halo and directly reflects the underlying Dehnen density distribution~\cite{Dehnen:1993uh,Taylor:2002zd},
\begin{equation}\label{Dehnendensity}
\rho(r)=\rho_{0}\left(\frac{r}{a}\right)^{-\alpha}\!\left(1+\frac{r^{k}}{a^{k}}\right)^{-(\gamma-\alpha)/k},
\end{equation}
with $\gamma=4$, $\alpha=0$ and $k=1$. At large radii, the spacetime smoothly approaches the Schwarzschild limit,
\begin{equation}
f(r)= 1-\frac{2M}{r}+\mathcal{O}\left(\frac{1}{r^2}\right),
\end{equation}
while at small $r$ the expansion
\begin{equation}
f(r)= 1-\frac{2Mr^{2}}{a^{3}}+\mathcal{O}(r^{3})
\end{equation}
reveals the presence of a regular de~Sitter core. Thus, the metric provides a minimal and physically motivated model of a black hole embedded in a realistic galactic environment, free from curvature singularities and exotic matter. In what follows, we adopt geometrized units $G=c=1$ and fix $M=1$, so that all quantities are expressed in units of the black-hole mass.

\begin{figure}
\resizebox{\linewidth}{!}{\includegraphics{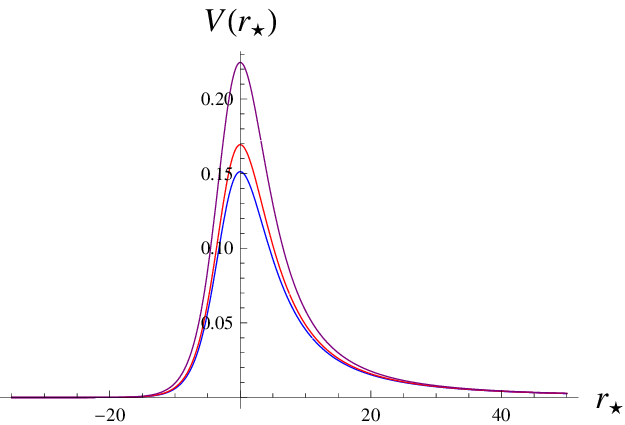}}
\caption{Effective potential as a function of the tortoise coordinate $r^{*}$ for $\ell=2$ up perturbations: $M=1$; $a=0$ (blue), $a=0.05$ (black) and $a=0.15$ (red).}\label{fig:upL2}
\end{figure}

\begin{figure}
\resizebox{\linewidth}{!}{\includegraphics{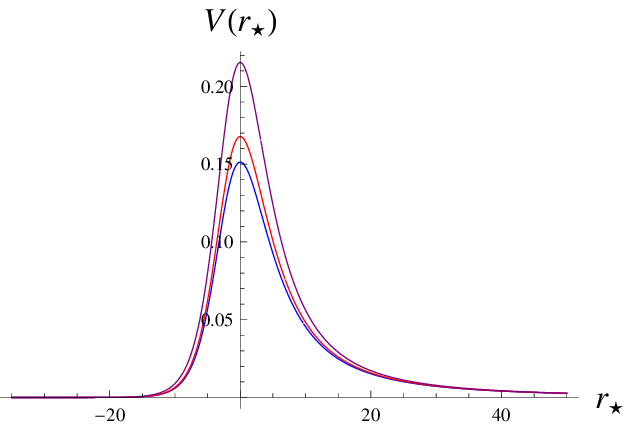}}
\caption{Effective potential as a function of the tortoise coordinate $r^{*}$ for $\ell=2$ down perturbations: $M=1$; $a=0$ (blue), $a=0.05$ (black) and $a=0.15$ (red).}\label{fig:downL2}
\end{figure}
Linear perturbations of spherically symmetric spacetimes supported by an anisotropic fluid require special care, because the background matter distribution introduces additional gauge-invariant degrees of freedom beyond those present in vacuum. A systematic derivation of the axial (odd-parity) sector for such configurations was presented in~\cite{Chakraborty:2024gcr}, where two distinct but equally consistent perturbative schemes were developed. These schemes differ in how variations of the fluid four–velocity and stress–energy components are constrained, and they ultimately lead to two inequivalent master equations. For convenience, we refer to these as the ``up'' and ``down'' perturbations.

\smallskip

To obtain the wave equations, one begins by expanding the metric perturbations in tensor spherical harmonics. Because the background spacetime is static and spherically symmetric, axial perturbations do not couple to scalar or vector matter perturbations at linear order. Consequently, after separating the angular dependence using spherical harmonics $Y_{\ell m}(\theta,\phi)$ and redefining the radial perturbation amplitudes through the standard Regge–Wheeler transformation, the dynamics reduce to a single Schrödinger-like equation for each multipole~$\ell$:
\begin{equation}
\frac{d^{2}\Psi}{dr_{*}^{2}}+\left(\omega^{2}-V(r)\right)\Psi=0,
\qquad 
dr_{*} \equiv \frac{dr}{f(r)},
\label{eq:master}
\end{equation}
where $\Psi(r)$ is the gauge-invariant master variable and $r_{*}$ is the tortoise coordinate, defined so as to map the exterior region onto $(-\infty,+\infty)$.

\smallskip

The distinction between the ``up'' and ``down'' formulations is rooted in the treatment of perturbations of the anisotropic fluid.  
In the ``up'' case, one works in a frame in which the \emph{contravariant} components of the fluid variables remain unperturbed, while in the ``down'' case, the perturbations of the \emph{covariant} components are suppressed instead. Although both approaches are physically legitimate and gauge-consistent, they lead to different effective potentials because the fluid stress tensor is anisotropic and therefore reacts differently to these constraints.

\smallskip

The two Regge--Wheeler–type effective potentials derived in~\cite{Chakraborty:2024gcr} take the form
\begin{eqnarray}
V^{\text{(up)}}(r) &=&
f(r)\Biggl[
\frac{\ell(\ell+1)}{r^{2}}
-\frac{6\,m(r)}{r^{3}}\\\nonumber&&
+4\pi\left(\rho(r)-5P_{r}(r)+4P(r)\right)
\Biggr], 
\label{Vup_general}
\\[2mm]
V^{\text{(down)}}(r) &=&
f(r)\Biggl[
\frac{\ell(\ell+1)}{r^{2}}
-\frac{6\,m(r)}{r^{3}}\\\nonumber&&
+4\pi\,\rho(r)-4\pi\,P_{r}(r)
\Biggr],
\label{Vdown_general}
\end{eqnarray}
where $m(r)$ is the Misner–Sharp mass function, while $\rho(r)$, $P_{r}(r)$ and $P(r)$ denote the energy density, radial pressure, and tangential pressure of the anisotropic fluid, respectively. The  nonzero components of the corresponding stress-energy tensor are
\begin{equation}\label{stress-energy}
\begin{array}{rcl}
T_{t}^{t} &=&-8\pi\rho(r),\\
T_{r}^{r} &=& ~~8\pi P_r(r), \\
T_{\theta}^{\theta} = T_{\varphi}^{\varphi} &=& ~~8\pi P(r),
\end{array}
\end{equation}
and the following conditions for the components of pressure are chosen in  \cite{Konoplya:2025ect}
\begin{eqnarray}\label{vacuum}
P_r(r)&=&-\rho(r),
\\\label{Pex}
P(r)&=&-\frac{r}{2}\rho'(r)-\rho(r).
\end{eqnarray}

Using the relations appropriate to our background configuration, in particular the conditions~(\ref{vacuum}) and~(\ref{Pex}), one may express the pressures entirely in terms of the density profile $\rho(r)$. This leads to a simplified pair of potentials:
\begin{eqnarray}
V^{\text{(up)}}(r) &=&
f(r)\Biggl[\frac{\ell(\ell+1)}{r^{2}}-\frac{6\,m(r)}{r^{3}} \\\nonumber&&
+8\pi\,\rho(r)-8\pi r\,\rho'(r)\Biggr], 
\label{Vup_simplified}
\\[2mm]
\!\!\!\!\!\!\!V^{\text{(down)}}(r) &=&
f(r)\Biggl[\frac{\ell(\ell+1)}{r^{2}}-\frac{6\,m(r)}{r^{3}}+8\pi\,\rho(r)\Biggr], \label{Vdown_simplified}
\end{eqnarray}
where a prime denotes differentiation with respect to~$r$.

Thus, although both perturbative schemes reduce to the familiar Regge–Wheeler potential in the vacuum limit $\rho(r)\to 0$, within an anisotropic-fluid halo they probe different combinations of the background matter variables. For a realistic astrophysical environment, this distinction becomes physically meaningful, and therefore both potentials must be analysed independently when computing quasinormal spectra, GBFs, and absorption cross-sections. The effective potentials are positive definite, as can be seen in Figs.~\ref{fig:upL2} and ~\ref{fig:downL2}, which guarantee the stability of perturbations and absence of unbounded modes growing in time. 

\section{Two methods for calculations of quasinormal modes}\label{sec:methodgsforQNMcal}

\subsection{WKB method}\label{sec:WKB}

When the master perturbation equation in the form
\[
\frac{d^{2}\Psi}{dr_{*}^{2}} + \bigl[\omega^{2} - V(r)\bigr]\Psi = 0,
\qquad dr_{*} = \frac{dr}{f(r)},
\]
is governed by an effective potential \(V(r)\) having the shape of a single barrier (with a maximum at \(r = r_{0}\)), the semi-analytic Wentzel–Kramers–Brillouin (WKB) method becomes applicable ~\cite{Iyer:1986np,Konoplya:2003ii,Matyjasek:2017psv}. Originally introduced by B. F. Schutz and C. M. Will and extended by subsequent authors, the WKB expansion provides an approximate expression for the complex frequencies \(\omega\) of QNMs (see examples in \cite{Bolokhov:2022rqv,Skvortsova:2024atk,Lutfuoglu:2025blw,del-Corral:2022kbk,Skvortsova:2023zmj,Zhao:2022gxl,Kodama:2009bf,Bolokhov:2025egl,Skvortsova:2024wly,Lutfuoglu:2025pzi,Lutfuoglu:2025ljm,Skvortsova:2025cah,Bolokhov:2025lnt,Skvortsova:2024msa,Dubinsky:2025wns,Zinhailo:2018ska}).

In its simplest (first) order, the WKB result corresponds to the eikonal (large-\(\ell\)) regime. However, for higher accuracy at finite multipole number \(\ell\), one introduces successive correction terms. In the form adopted in the literature, the squared frequency is expanded as:
\begin{eqnarray}\label{eq:WKBbasic}
\omega^{2} &=& V_{0} + \Lambda_{2} + \Lambda_{4} + \Lambda_{6} + \dots\\\nonumber&&
- i\,\left(n + \tfrac{1}{2}\right)\sqrt{-2\,V_{2}}\,\Bigl( 1 + \Lambda_{3} + \Lambda_{5} + \dots \Bigr),
\end{eqnarray}
where \(n=0,1,2,\dots\) is the overtone number, and
\[V_{i} = \left.\frac{d^{i}V}{dr_{*}^{i}}\right|_{r_{*}(r_{0})}\]
denotes derivatives of the effective potential at its peak. The terms \(\Lambda_{k}\) are WKB correction coefficients depending on \(V_{0},V_{2},V_{3},\dots,V_{2k}\).

As demonstrated in the reference work, the conventional WKB series is asymptotic and may not converge uniformly for all potentials. To enhance reliability and extend the domain of applicability, the use of Padé approximants is recommended. Specifically, one rewrites the WKB expansion for \(\omega^{2}\) (or \(\omega\)) as a rational function:
\begin{equation}\label{eq:PadeWKB}
P_{k/m}\{\omega^{2}\} \;=\; \frac{\sum_{i=0}^{k} a_{i}\,(\omega^{2})^{i}}{1 + \sum_{j=1}^{m} b_{j}\,(\omega^{2})^{j}}
\end{equation}
where the indices \(k\) and \(m\) denote the degrees of the numerator and denominator polynomials chosen so as to best fit the truncated WKB expansion. The resulting expression produces much improved accuracy with typically smaller deviations when compared to reference “exact” numerical methods.

In practice, comparing Padé approximants of different orders with $k\approx m$ and examining the spread of results is used to estimate the error of the approximation. The paper provides “recipes” for error estimation and discusses in which situations the WKB+Padé method remains reliable.

In the present study, we apply the sixth- and ninth-order (WKB\(_6\) and WKB\(_9\)) formulas in conjunction with Padé approximants. Our baseline approach is as follows:
\begin{itemize}
  \item Compute \(V_{0},V_{2},\dots\) up to the appropriate order at the peak of the effective potential \(r_{0}\).
  \item Evaluate \(\omega\) from equation~\eqref{eq:WKBbasic} using both orders.
  \item Apply a suitable Padé approximation (e.g., \(P_{3/3}\) or \(P_{4/5}\)) and determine the final frequency \(\omega\).
  \item Assess convergence by comparing WKB\(_6\) and WKB\(_9\) results: the relative percentage difference is reported in the tables of Section~\ref{sec:QNMs}.
\end{itemize}

This procedure yields accurate quasinormal frequencies for scalar, electromagnetic, Dirac and—most importantly in this work—gravitational perturbations of the regular black-hole metric under consideration, as numerous examples demonstrate \cite{Bolokhov:2023bwm, Konoplya:2023ahd, Bolokhov:2024ixe, Albuquerque:2023lhm, Lutfuoglu:2025eik, Ishihara:2008re, Bonanno:2025dry, Guo:2022hjp, Zhidenko:2003wq, Skvortsova:2024eqi, Konoplya:2022hbl, Bolokhov:2023ruj, Abdalla:2005hu, Paul:2023eep, Arbelaez:2025gwj, Lutfuoglu:2025qkt, Konoplya:2001ji, Kokkotas:2010zd, Lutfuoglu:2025bsf, Zinhailo:2019rwd}. The deviation between successive orders is used as an internal consistency check; agreement to within one part in \(10^{-3}\) (or better) supports the robustness of our results.

While the WKB+Padé method demonstrates strong performance for single-barrier potentials with a dominant peak and moderate overtone number (\(n < \ell\)), it has known limitations:
\begin{itemize}
  \item For potentials with multiple peaks or strong asymptotic oscillatory tails, the WKB series may diverge or lose accuracy.
  \item Extremely high overtones or low multipole numbers may deviate significantly from the eikonal regime, reducing reliability.
  \item Care must be taken when applying the method to potentials that do not tend to constant values at the boundaries (e.g., pure AdS or near-extremal cases without a clean barrier).
\end{itemize}
As shown in the reference and verified in our computations, the WKB+Padé method works very well for the family of metrics analysed here.

\subsection{Time-domain evolution}\label{subsec:timedomain}

An important cross–check of the WKB-based quasinormal spectrum is provided by direct integration of the perturbation equation in the time domain. To this end, we evolve the master wave equation,
\begin{equation}
\label{eq:wave-time}
\frac{\partial^{2}\Psi(t,r_{*})}{\partial t^{2}}
 - \frac{\partial^{2}\Psi(t,r_{*})}{\partial r_{*}^{2}}
 + V(r)\,\Psi(t,r_{*}) = 0,
\end{equation}
on a discretized characteristic grid of the light-cone coordinates \(u = t - r_{*}\) and \(v = t + r_{*}\). We employ the standard second-order convergent Gundlach--Price--Pullin finite-difference scheme \cite{Gundlach:1993tp}, which updates the field at each grid point according to
\begin{equation}
\Psi_{N} = \Psi_{W} + \Psi_{E} - \Psi_{S} - \frac{\Delta^{2}}{8}\,V_{S}\left(\Psi_{W} + \Psi_{E}\right) + \mathcal{O}(\Delta^{4}),
\end{equation}
where the subscripts \(N, S, E, W\) refer to the values of the field at the future, past, and spatially adjacent points of a null rectangle.

Initial data are specified on the two initial null segments, typically in the form of a Gaussian pulse imposed on \(u=u_{0}\),
\begin{equation}
\Psi(u_{0},v) = \exp\!\left[-\frac{(v - v_{c})^{2}}{2\sigma^{2}}\right],
\end{equation}
with \(\Psi(v_{0},u)=const\) on the other leg of the initial surface. The signal extracted at fixed \(r_{*}\) exhibits the familiar sequence of prompt response, quasinormal ringing, and late-time tails.   The quasinormal frequencies are obtained by applying the Prony method to the exponentially damped oscillations in the intermediate regime.

Agreement between the WKB predictions and the frequencies extracted from the time-domain signal provides a robust consistency check \cite{Malik:2024nhy,Konoplya:2007jv,Skvortsova:2023zca,Aneesh:2018hlp,Dubinsky:2024mwd,Konoplya:2024lch,Qian:2022kaq,Momennia:2022tug,Dubinsky:2024gwo,Malik:2024tuf,Konoplya:2025uiq,Cuyubamba:2016cug,Bronnikov:2021liv,Malik:2024elk,Konoplya:2013sba,Dubinsky:2024aeu,Konoplya:2025hgp,Malik:2024qsz,Bolokhov:2024bke}.  In all cases studied here---for both ``up'' and ``down'' gravitational sectors and for all values of the halo parameter \(a\)---the time-domain evolution confirms the WKB results to the accuracy expected for low overtones.

\begin{figure*}
\resizebox{\linewidth}{!}{\includegraphics{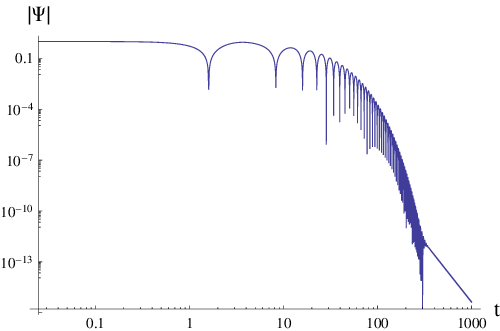}\includegraphics{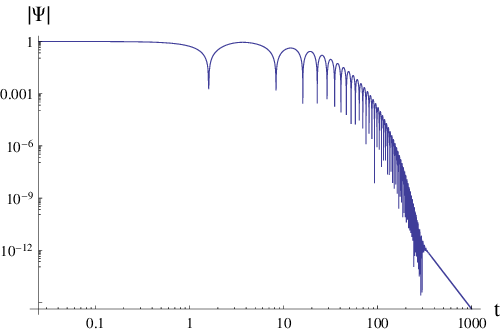}}
\caption{The logarithmic time-domain profile for $\ell=2$ up-perturbations (left) and down-perturbations (right) at $a=0.25$, $M=1$. The fundamental QNMs given by the Prony extraction method are $\omega = 0.58167 - 0.09089 i$ (for up - modes) and $\omega = 0.54474 - 0.09191 i$ (for down - modes), which coincide with the WKB data. The asymptotic tail is $\sim t^{-7}$ in both cases, which coincides with the Price law for the Schwarzschild spacetime.} \label{fig:TD}
\end{figure*}

\begin{table}
\centering
\footnotesize
\begin{tabular}{l c c l}
\hline
$a$ & WKB6 ($m=3$) & WKB9 ($m=5$) & rel. diff.  \\
\hline
$0$ & $0.3736199-0.0889328 i$ & $0.3736588-0.0889734 i$ & $0.0146\%$\\
$0.02$ & $0.3824583-0.0894966 i$ & $0.3824643-0.0895187 i$ & $0.00583\%$\\
$0.04$ & $0.3919160-0.0900504 i$ & $0.3919188-0.0900619 i$ & $0.00294\%$\\
$0.06$ & $0.4021189-0.0906065 i$ & $0.4021155-0.0905975 i$ & $0.00235\%$\\
$0.08$ & $0.4131646-0.0911225 i$ & $0.4131693-0.0911161 i$ & $0.00187\%$\\
$0.1$ & $0.4252182-0.0916091 i$ & $0.4252238-0.0916090 i$ & $0.00130\%$\\
$0.12$ & $0.4384554-0.0920588 i$ & $0.4384599-0.0920627 i$ & $0.00134\%$\\
$0.14$ & $0.4531063-0.0924561 i$ & $0.4531096-0.0924593 i$ & $0.000975\%$\\
$0.16$ & $0.4694507-0.0927632 i$ & $0.4694835-0.0927696 i$ & $0.00700\%$\\
$0.18$ & $0.4879955-0.0929517 i$ & $0.4879974-0.0929539 i$ & $0.000580\%$\\
$0.2$ & $0.5092505-0.0929461 i$ & $0.5092537-0.0929388 i$ & $0.00154\%$\\
$0.22$ & $0.5341402-0.0926097 i$ & $0.5341423-0.0926011 i$ & $0.00163\%$\\
$0.24$ & $0.5640334-0.0916870 i$ & $0.5640892-0.0916986 i$ & $0.00996\%$\\
$0.26$ & $0.6015909-0.0897065 i$ & $0.6015910-0.0897056 i$ & $0.000150\%$\\
$0.28$ & $0.6518269-0.0851445 i$ & $0.6518313-0.0850873 i$ & $0.00873\%$\\
\hline
\end{tabular}
\caption{Quasinormal modes of the $\ell=2$ up–potential for the Konoplya–Zhidenko black hole ($M=1$), computed with the 6th–order and 9th–order WKB methods with Padé approximants. The relative difference between the two WKB–Padé approximants is given in percent.}
\end{table}

\begin{table}
\centering
\footnotesize
\begin{tabular}{l c c l}
\hline
$a$ & WKB6 ($m=3$) & WKB9 ($m=5$) & rel. diff.  \\
\hline
$0$ & $0.5994434-0.0927029 i$ & $0.5994433-0.0927031 i$ & $0.000029\%$\\
$0.02$ & $0.6126050-0.0933460 i$ & $0.6126051-0.0933459 i$ & $0.000022\%$\\
$0.04$ & $0.6266776-0.0939827 i$ & $0.6266779-0.0939826 i$ & $0.000050\%$\\
$0.06$ & $0.6417829-0.0946073 i$ & $0.6417833-0.0946072 i$ & $0.000067\%$\\
$0.08$ & $0.6580683-0.0952119 i$ & $0.6580688-0.0952118 i$ & $0.000076\%$\\
$0.1$ & $0.6757154-0.0957853 i$ & $0.6757159-0.0957852 i$ & $0.000074\%$\\
$0.12$ & $0.6949507-0.0963117 i$ & $0.6949511-0.0963117 i$ & $0.000055\%$\\
$0.14$ & $0.7160632-0.0967685 i$ & $0.7160632-0.0967685 i$ & $8.6\times 10^{-6}\%$\\
$0.16$ & $0.7394288-0.0971213 i$ & $0.7394282-0.0971223 i$ & $0.000153\%$\\
$0.18$ & $0.7655488-0.0973199 i$ & $0.7655498-0.0973217 i$ & $0.000265\%$\\
$0.2$ & $0.7951253-0.0972840 i$ & $0.7951266-0.0972848 i$ & $0.000191\%$\\
$0.22$ & $0.8291711-0.0968724 i$ & $0.8291722-0.0968726 i$ & $0.000132\%$\\
$0.24$ & $0.8692489-0.0958295 i$ & $0.8692497-0.0958296 i$ & $0.000087\%$\\
$0.26$ & $0.9179902-0.0936226 i$ & $0.9179906-0.0936225 i$ & $0.000049\%$\\
$0.28$ & $0.9804855-0.0888796 i$ & $0.9804857-0.0888795 i$ & $0.000019\%$\\
\hline
\end{tabular}
\caption{Quasinormal modes of the $\ell=3$ up–potential for the Konoplya–Zhidenko black hole ($M=1$), computed with the 6th–order and 9th–order WKB methods with Padé approximants. The relative difference between the two WKB–Padé approximants is given in percent.}
\end{table}

\begin{table}
\centering
\footnotesize
\begin{tabular}{l c c l}
\hline
$a$ & WKB6 ($m=3$) & WKB9 ($m=5$) & rel. diff.  \\
\hline
$0$ & $0.3736199-0.0889328 i$ & $0.3736588-0.0889734 i$ & $0.0146\%$\\
$0.02$ & $0.3816645-0.0895694 i$ & $0.3816841-0.0896053 i$ & $0.0104\%$\\
$0.04$ & $0.3902288-0.0902091 i$ & $0.3902369-0.0902341 i$ & $0.00655\%$\\
$0.06$ & $0.3993786-0.0908392 i$ & $0.3993835-0.0908542 i$ & $0.00387\%$\\
$0.08$ & $0.4091983-0.0914469 i$ & $0.4092042-0.0914583 i$ & $0.00307\%$\\
$0.1$ & $0.4197974-0.0920216 i$ & $0.4197969-0.0920366 i$ & $0.00348\%$\\
$0.12$ & $0.4312904-0.0925722 i$ & $0.4312829-0.0925749 i$ & $0.00181\%$\\
$0.14$ & $0.4438166-0.0930546 i$ & $0.4438148-0.0930530 i$ & $0.000546\%$\\
$0.16$ & $0.4575866-0.0934419 i$ & $0.4575881-0.0934410 i$ & $0.000379\%$\\
$0.18$ & $0.4728569-0.0936933 i$ & $0.4728603-0.0936939 i$ & $0.000723\%$\\
$0.2$ & $0.4899763-0.0937387 i$ & $0.4899806-0.0937406 i$ & $0.000941\%$\\
$0.22$ & $0.5094378-0.0934597 i$ & $0.5094420-0.0934615 i$ & $0.000888\%$\\
$0.24$ & $0.5319761-0.0926420 i$ & $0.5319778-0.0926424 i$ & $0.000334\%$\\
$0.26$ & $0.5587646-0.0908387 i$ & $0.5587577-0.0908493 i$ & $0.00223\%$\\
$0.28$ & $0.5918319-0.0870297 i$ & $0.5918373-0.0870358 i$ & $0.00136\%$\\
\hline
\end{tabular}
\caption{Quasinormal modes of the $\ell=2$ down–potential for the Konoplya–Zhidenko black hole ($M=1$), computed with the 6th–order and 9th–order WKB methods with Padé approximants. The relative difference between the two WKB–Padé approximants is given in percent.}
\end{table}

\begin{table}
\centering
\footnotesize
\begin{tabular}{l c c l}
\hline
$a$ & WKB6 ($m=3$) & WKB9 ($m=5$) & rel. diff.  \\
\hline
$0$ & $0.5994434-0.0927029 i$ & $0.5994433-0.0927031 i$ & $0.000029\%$\\
$0.02$ & $0.6120647-0.0933597 i$ & $0.6120646-0.0933598 i$ & $0.000012\%$\\
$0.04$ & $0.6255137-0.0940110 i$ & $0.6255137-0.0940111 i$ & $1.5\times 10^{-6}\%$\\
$0.06$ & $0.6398952-0.0946512 i$ & $0.6398952-0.0946511 i$ & $7.4\times 10^{-6}\%$\\
$0.08$ & $0.6553349-0.0952724 i$ & $0.6553350-0.0952723 i$ & $0.000014\%$\\
$0.1$ & $0.6719861-0.0958639 i$ & $0.6719862-0.0958638 i$ & $0.000018\%$\\
$0.12$ & $0.6900380-0.0964107 i$ & $0.6900381-0.0964106 i$ & $0.000020\%$\\
$0.14$ & $0.7097286-0.0968912 i$ & $0.7097287-0.0968911 i$ & $0.000021\%$\\
$0.16$ & $0.7313634-0.0972739 i$ & $0.7313636-0.0972738 i$ & $0.000019\%$\\
$0.18$ & $0.7553447-0.0975109 i$ & $0.7553448-0.0975108 i$ & $0.000015\%$\\
$0.2$ & $0.7822185-0.0975271 i$ & $0.7822186-0.0975270 i$ & $9.2\times 10^{-6}\%$\\
$0.22$ & $0.8127574-0.0971974 i$ & $0.8127574-0.0971973 i$ & $3.0\times 10^{-6}\%$\\
$0.24$ & $0.8481131-0.0962975 i$ & $0.8481130-0.0962975 i$ & $9.3\times 10^{-6}\%$\\
$0.26$ & $0.8901372-0.0943783 i$ & $0.8901371-0.0943784 i$ & $0.000022\%$\\
$0.28$ & $0.9421602-0.0903623 i$ & $0.9421600-0.0903627 i$ & $0.000051\%$\\
\hline
\end{tabular}
\caption{Quasinormal modes of the $\ell=3$ down–potential for the Konoplya–Zhidenko black hole ($M=1$), computed with the 6th–order and 9th–order WKB methods with Padé approximants. The relative difference between the two WKB–Padé approximants is given in percent.}
\end{table}

\begin{table}
\centering
\footnotesize
\begin{tabular}{r c c l}
\hline
$\ell$ & WKB6 ($m=3$) & WKB9 ($m=5$) & rel. diff.  \\
\hline
$2$ & $0.53414017-0.09260969 i$ & $0.53414233-0.09260111 i$ & $0.00163\%$\\
$3$ & $0.82917113-0.09687236 i$ & $0.82917219-0.09687265 i$ & $0.000132\%$\\
$4$ & $1.10688988-0.09828610 i$ & $1.10688981-0.09828629 i$ & $0.0000184\%$\\
$5$ & $1.37754858-0.09893886 i$ & $1.37754857-0.09893888 i$ & $1.6\times 10^{-6}\%$\\
$10$ & $2.69784668-0.09982890 i$ & $2.69784668-0.09982890 i$ & $0\%$\\
$15$ & $4.00280323-0.10000156 i$ & $4.00280323-0.10000156 i$ & $0\%$\\
$20$ & $5.30373817-0.10006351 i$ & $5.30373817-0.10006351 i$ & $0\%$\\
$25$ & $6.60302732-0.10009263 i$ & $6.60302732-0.10009263 i$ & $0\%$\\
$30$ & $7.90148225-0.10010860 i$ & $7.90148225-0.10010860 i$ & $0\%$\\
\hline
\end{tabular}
\caption{Quasinormal modes for the up–potential with $a=0.22$ for the Konoplya–Zhidenko black hole ($M=1$), computed with the 6th–order and 9th–order WKB methods with Padé approximants, as a function of multipole number $\ell$. The relative difference between the two WKB–Padé approximants is given in percent.}
\end{table}

\begin{table}
\centering
\footnotesize
\begin{tabular}{r c c l}
\hline
$\ell$ & WKB6 ($m=3$) & WKB9 ($m=5$) & rel. diff.  \\
\hline
$2$ & $0.50943778-0.09345968 i$ & $0.50944202-0.09346146 i$ & $0.000888\%$\\
$3$ & $0.81275737-0.09719736 i$ & $0.81275736-0.09719733 i$ & $3\times 10^{-6}\%$\\
$4$ & $1.09449251-0.09847279 i$ & $1.09449250-0.09847280 i$ & $4.3\times 10^{-7}\%$\\
$5$ & $1.36755327-0.09906164 i$ & $1.36755326-0.09906164 i$ & $2.2\times 10^{-7}\%$\\
$10$ & $2.69272139-0.09986200 i$ & $2.69272139-0.09986200 i$ & $0\%$\\
$15$ & $3.99934619-0.10001671 i$ & $3.99934619-0.10001671 i$ & $0\%$\\
$20$ & $5.30112839-0.10007217 i$ & $5.30112839-0.10007217 i$ & $0\%$\\
$25$ & $6.60093080-0.10009822 i$ & $6.60093080-0.10009822 i$ & $0\%$\\
$30$ & $7.89973014-0.10011251 i$ & $7.89973014-0.10011251 i$ & $0\%$\\
\hline
\end{tabular}
\caption{Quasinormal modes for the down–potential with $a=0.22$ for the Konoplya–Zhidenko black hole ($M=1$), computed with the 6th–order and 9th–order WKB methods with Padé approximants, as a function of multipole number $\ell$. The relative difference between the two WKB–Padé approximants is given in percent.}
\end{table}

\begin{table}
\centering
\footnotesize
\begin{tabular}{l c c lc}
\hline
$a$ & WKB6 ($m=3$) & WKB9 ($m=5$) & rel. diff  \\
\hline.
$0$ & $0.3460075-0.2735658 i$ & $0.3474780-0.2742904 i$ & $0.372\%$\\
$0.02$ & $0.3560721-0.2750344 i$ & $0.3567270-0.2754235 i$ & $0.169\%$\\
$0.04$ & $0.3668714-0.2766587 i$ & $0.3670814-0.2769795 i$ & $0.0834\%$\\
$0.06$ & $0.3782626-0.2784519 i$ & $0.3781467-0.2784559 i$ & $0.0247\%$\\
$0.08$ & $0.3902395-0.2800427 i$ & $0.3904275-0.2801449 i$ & $0.0446\%$\\
$0.1$ & $0.4032950-0.2812501 i$ & $0.4039825-0.2812245 i$ & $0.140\%$\\
$0.12$ & $0.4177546-0.2822442 i$ & $0.4184016-0.2821639 i$ & $0.129\%$\\
$0.14$ & $0.4337959-0.2830781 i$ & $0.4343173-0.2830092 i$ & $0.102\%$\\
$0.16$ & $0.4516638-0.2836631 i$ & $0.4521399-0.2836757 i$ & $0.0893\%$\\
$0.18$ & $0.4717668-0.2838653 i$ & $0.4719549-0.2838005 i$ & $0.0361\%$\\
$0.2$ & $0.4946774-0.2833980 i$ & $0.4949448-0.2833510 i$ & $0.0476\%$\\
$0.22$ & $0.5212659-0.2817904 i$ & $0.5214530-0.2817233 i$ & $0.0335\%$\\
$0.24$ & $0.5528904-0.2784259 i$ & $0.5529053-0.2781922 i$ & $0.0378\%$\\
$0.26$ & $0.5918902-0.2714048 i$ & $0.5919437-0.2713890 i$ & $0.00856\%$\\
$0.28$ & $0.6432812-0.2571667 i$ & $0.6428211-0.2561781 i$ & $0.157\%$\\
\hline
\end{tabular}
\caption{Quasinormal modes of the $\ell=2$, $n=1$ up–potential for the Konoplya–Zhidenko black hole ($M=1$), computed with the 6th–order and 9th–order WKB methods with Padé approximants. The relative difference between the two WKB–Padé approximants is given in percent.}
\end{table}

\begin{table}
\centering
\footnotesize
\begin{tabular}{l c c l}
\hline
$a$ & WKB6 ($m=3$) & WKB9 ($m=5$) & rel. diff.  \\
\hline
$0$ & $0.3460075-0.2735658 i$ & $0.3474780-0.2742904 i$ & $0.372\%$\\
$0.02$ & $0.3548282-0.2753178 i$ & $0.3558712-0.2757730 i$ & $0.253\%$\\
$0.04$ & $0.3642295-0.2770634 i$ & $0.3649320-0.2774246 i$ & $0.173\%$\\
$0.06$ & $0.3742778-0.2787919 i$ & $0.3747251-0.2791016 i$ & $0.117\%$\\
$0.08$ & $0.3850476-0.2804864 i$ & $0.3853037-0.2807361 i$ & $0.0751\%$\\
$0.1$ & $0.3966244-0.2821139 i$ & $0.3967538-0.2822929 i$ & $0.0454\%$\\
$0.12$ & $0.4091168-0.2836130 i$ & $0.4091062-0.2836971 i$ & $0.0170\%$\\
$0.14$ & $0.4226812-0.2848877 i$ & $0.4226583-0.2848694 i$ & $0.00576\%$\\
$0.16$ & $0.4375432-0.2858156 i$ & $0.4374374-0.2855834 i$ & $0.0488\%$\\
$0.18$ & $0.4540024-0.2862479 i$ & $0.4540708-0.2861105 i$ & $0.0286\%$\\
$0.2$ & $0.4724397-0.2859715 i$ & $0.4725108-0.2858708 i$ & $0.0223\%$\\
$0.22$ & $0.4933573-0.2846242 i$ & $0.4934057-0.2845615 i$ & $0.0139\%$\\
$0.24$ & $0.5174526-0.2815178 i$ & $0.5174488-0.2815083 i$ & $0.00174\%$\\
$0.26$ & $0.5455665-0.2752612 i$ & $0.5456136-0.2753838 i$ & $0.0215\%$\\
$0.28$ & $0.5789902-0.2630174 i$ & $0.5790632-0.2630410 i$ & $0.0121\%$\\
\hline
\end{tabular}
\caption{Quasinormal modes of the $\ell=2$, $n=1$ down–potential for the Konoplya–Zhidenko black hole ($M=1$), computed with the 6th–order and 9th–order WKB methods with Padé approximants. The relative difference between the two WKB–Padé approximants is given in percent.}
\end{table}

\section{Quasinormal modes}\label{sec:QNMs}

The numerical analysis of gravitational perturbations of the Konoplya-Zhidenko regular black hole supported by a Dehnen-type matter profile reveals several robust and physically meaningful trends. The dominant dependence is controlled by the halo scale parameter \(a\), which quantifies the effective density and spatial extent of the surrounding dark-matter distribution. For both axial ``up'' and ``down'' perturbations, increasing \(a\) leads to a monotonic rise in the real part of the quasinormal frequencies, \(\re{\omega}\), accompanied by only a very mild modification of the damping rate (see Tables I-IV). This behavior reflects the fact that enlarging \(a\) slightly strengthens the effective potential barrier and shifts its peak outward, resulting in higher oscillation frequencies while leaving the imaginary parts \(\im{\omega}\) nearly unchanged. Even for the largest values considered (\(a \simeq 0.30\)), the change in \(\im{\omega}\) remains below a few parts in \(10^{-3}\), showing that the halo affects the ringdown longevity only weakly.

The dependence on multipole number \(\ell\) follows the expected trend: higher-\(\ell\) modes lie deeper in the eikonal regime and are therefore less sensitive to the dark-matter environment (see Tables V and VI). While the shift in \\\(\re{\omega}\) due to the halo is clearly visible for \(\ell=2\), it quickly becomes negligible as \(\ell\) increases, consistent with the centrifugal barrier dominating the effective potential at large \(\ell\). This is confirmed by the excellent agreement between WKB predictions of different orders for \(\ell\ge 3\), where the relative difference between sixth-order WKB with Padé \(P_{3/3}\) and ninth-order with Padé \(P_{4/5}\) drops to \(10^{-5}\)--\(10^{-7}\), far smaller than the physical effect induced by the halo. Even for the fundamental quadrupole mode, the numerical convergence is strong: the difference between WKB6 and WKB9 is typically an order of magnitude smaller than the variation of \(\re{\omega}\) due to changing \(a\), demonstrating that the numerical accuracy of the method is fully sufficient to resolve the halo-induced shifts.

A distinctive feature of perturbations in this system is the breaking of isospectrality between the ``up'' and ``down'' axial sectors, as shown in Tables I-VIII. While in vacuum General Relativity odd and even gravitational perturbations share the same spectrum, here the anisotropic fluid profile affects the perturbation equations differently depending on whether covariant or contravariant matter components are varied within the same axial channel of perturbations. As a result, the two potentials produce slightly different quasinormal frequencies. The deviation is small but systematic: for fixed \(a\) and \(\ell\), the ``up'' modes consistently show marginally higher values of \(\re{\omega}\) compared to the ``down'' ones. This splitting grows with \(a\), reflecting the increasingly non-vacuum character of the background, but remains at the percent level or below for all considered cases. The time-domain profiles shown in Fig.~\ref{fig:TD}  further confirm the WKB results and validate the broken isospectrality: the extracted frequencies match the WKB predictions to within the expected numerical accuracy, with no indication of late-time numerical instabilities. 

Overall, the numerical data demonstrate that the quasinormal spectrum of this regular dark-matter-supported black hole is stable, weakly damped, and only moderately sensitive to the halo parameter \(a\). The deviations from the Schwarzschild spectrum are dominated by shifts in the oscillation frequency, while the damping rates remain essentially unchanged. The broken isospectrality provides an additional characteristic signature of the matter-supported geometry, though the effect is too small to be observationally relevant at present. The internal consistency between WKB and time-domain methods, together with the proximity of the results obtained by WKB method at different orders, confirms that the obtained spectra provide a reliable and accurate description of the gravitational ringdown in this model.

Using the data presented in Tables~V and VI, we observe that for sufficiently large $\ell$ the eikonal formula derived in~\cite{Bolokhov:2025fto} 
\begin{eqnarray}\nonumber
&&
\omega = \left(\ell+\frac{1}{2}\right) \left(\frac{1}{3 \sqrt{3} M}+\frac{a}{3 \sqrt{3}
   M^2}+\frac{a^2}{2 \sqrt{3}
   M^3}\right)
   \\\nonumber&&\!\!\!\!\!\!
   +i \left(n+\frac{1}{2}\right) \left(-\frac{1}{3
   \sqrt{3} M}-\frac{a}{9 \sqrt{3} M^2}+\frac{a^2}{54 \sqrt{3}
   M^3}\right).
\end{eqnarray}
is accurately reproduced. Following the general expansion techniques in inverse powers of $\ell$ developed in~\cite{Konoplya:2020hyk,Konoplya:2023moy,Dubinsky:2024rvf}, one could, in principle, obtain higher-order corrections beyond the eikonal limit. However, the resulting expressions become increasingly cumbersome, so we do not present them here.

At asymptotically late times, as illustrated in Fig.~\ref{fig:TD}, the evolution follows the Price law  \cite{Price:1972pw} for power-law tails,
\[
|\Psi| \sim t^{-(2\ell + 3)}, \qquad t \rightarrow \infty ,
\]
so that at sufficiently late times the waveform becomes indistinguishable from that of the vacuum Schwarzschild black hole, regardless of the presence of the halo.

\begin{figure}
\resizebox{\linewidth}{!}{\includegraphics{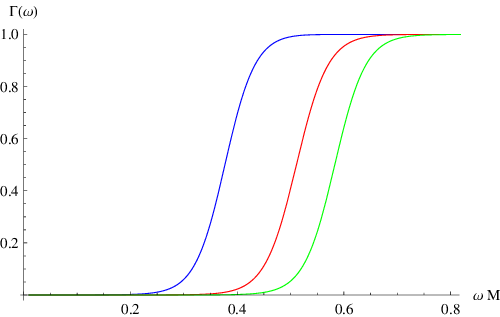}}
\caption{Grey-body factors for $\ell=2$ up perturbations at $a=0$ (blue), $a=0.2$ (red) and $a=0.25$ (green).}\label{fig:GBFupL2}
\end{figure}

\begin{figure}
\resizebox{\linewidth}{!}{\includegraphics{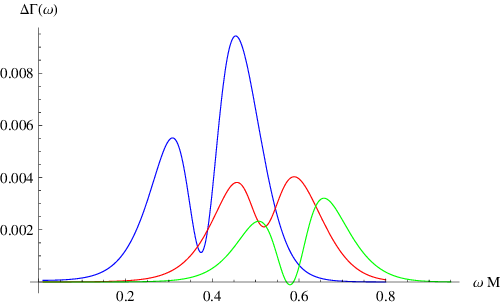}}
\caption{Difference between GBFs obtained by the 6th order WKB formula and the correspondence for $\ell=2$ up perturbations at $a=0$ (blue), $a=0.2$ (red) and $a=0.25$ (green).}\label{fig:GBFupL2DIFF}
\end{figure}

\begin{figure}
\resizebox{\linewidth}{!}{\includegraphics{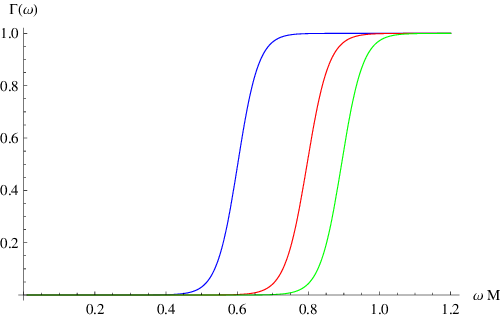}}
\caption{Grey-body factors for $\ell=3$ up - perturbations at $a=0$ (blue), $a=0.2$ (red) and $a=0.25$ (green).}\label{fig:GBFupL3}
\end{figure}

\begin{figure}
\resizebox{\linewidth}{!}{\includegraphics{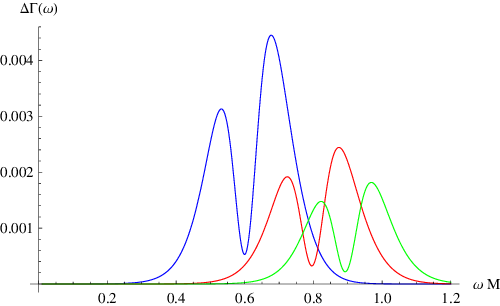}}
\caption{Difference between GBFs obtained by the 6th order WKB formula and the correspondence for $\ell=3$ up - perturbations at $a=0$ (blue), $a=0.2$ (red) and $a=0.25$ (green).}\label{fig:GBFupL3DIFF}
\end{figure}

\begin{figure}
\resizebox{\linewidth}{!}{\includegraphics{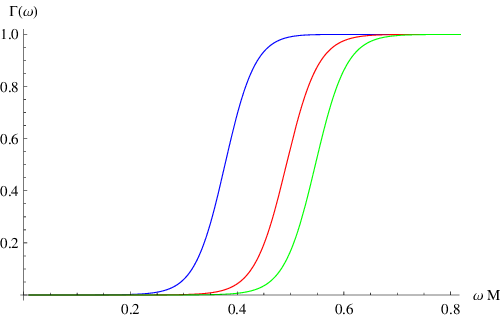}}
\caption{Grey-body factors for $\ell=2$ down - perturbations at $a=0$ (blue), $a=0.2$ (red) and $a=0.25$ (green).}\label{fig:GBFdownL3}
\end{figure}

\begin{figure}
\resizebox{\linewidth}{!}{\includegraphics{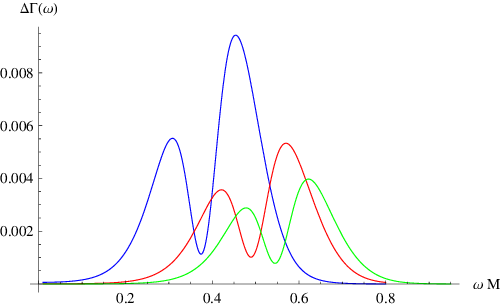}}
\caption{Difference between GBFs obtained by the 6th order WKB formula and the correspondence for $\ell=2$ down - perturbations at $a=0$ (blue), $a=0.2$ (red) and $a=0.25$ (green).}\label{fig:GBFdownL3DIFF}
\end{figure}

\begin{figure*}
\resizebox{\linewidth}{!}{\includegraphics{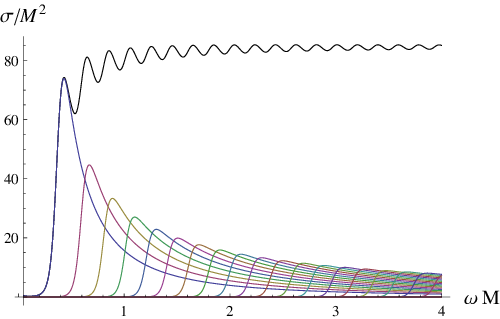}\includegraphics{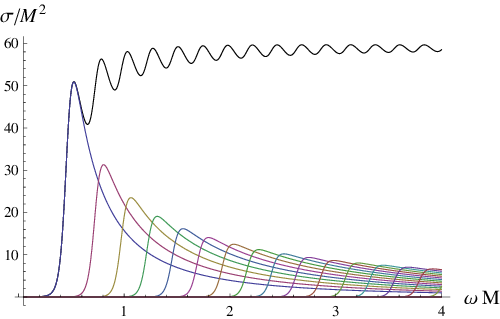}\includegraphics{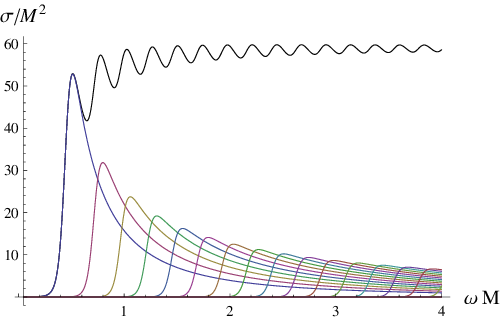}}
\caption{Absorption cross section for individual $\ell$ and the sum over the first 50 multipole numbers for up-perturbations; $M=1$, $a=0$ (left) and $a=0.15$ up-perturbations (middle) and $a=0.15$ down-perturbations (right). The presence of a nonzero halo parameter strongly suppresses the absorption cross-sections. The difference between cross-sections for up and down perturbations is very small.}\label{fig:SIGMA}
\end{figure*}

\section{Grey-body factors and absorption cross-sections}\label{sec:cross-sections}

Because the potential forms a barrier that decays both at the horizon and at spatial infinity, the scattering problem is defined by imposing the physically appropriate boundary conditions:
\begin{eqnarray}
\Psi(r_{*}) &\sim& e^{-i\omega r_{*}}, 
\qquad r_{*}\to -\infty 
\nonumber\\&&\text{(purely ingoing at the horizon)}, 
\\[2mm]
\Psi(r_{*}) &\sim& 
A_{\rm out}\, e^{+i\omega r_{*}}
+ A_{\rm in}\, e^{-i\omega r_{*}}, 
\qquad r_{*}\to +\infty 
\nonumber\\&&\text{(incident plus reflected waves)} .
\end{eqnarray}
The coefficients $A_{\rm in}$ and $A_{\rm out}$ determine,
respectively, the amplitudes of the incoming and outgoing waves at
spatial infinity.  
The grey-body factor (or transmission probability) is defined as
\begin{equation}
\Gamma_{\ell}(\omega)
   = 1 - \left|\frac{A_{\rm out}}{A_{\rm in}}\right|^{2}
   = \frac{|T_{\ell}(\omega)|^{2}}{|A_{\rm in}|^{2}} ,
\end{equation}
where $T_{\ell}(\omega)$ is the transmission amplitude through the effective potential barrier. It measures the fraction of the wave that reaches the horizon and, therefore, quantifies the deviation of black-hole radiation from a perfect black body.

\medskip

Grey-body factors can be computed using the same WKB framework employed for quasinormal mode calculations, but in this case without Padé resummation. The WKB expression for the transmission probability takes the compact form
\begin{equation}\label{eq:gbfactor}
\Gamma_{\ell}(\omega) = \frac{1}{\,1 + e^{2\pi i \mathcal{K}}\,},
\end{equation}
where the function \( \mathcal{K} \) depends on the local structure of the effective potential barrier and its derivatives up to order \( 2t \) in the \( t \)-th order WKB expansion~\cite{
Schutz:1985km,
Iyer:1986np,
Konoplya:2003ii,
Matyjasek:2017psv}.
This WKB method has been widely applied to compute transmission coefficients for black holes and wormholes in a variety of contexts~\cite{
Konoplya:2010kv,
Dubinsky:2024nzo,
Fernando:2016ksb,
Konoplya:2023ahd,Konoplya:2009hv,Lutfuoglu:2025ldc,Lutfuoglu:2025ohb,Bunjusuwan:2025enh}.

\medskip

The QNM spectrum is closely related to the resonant structure of the GBFs. Near a quasinormal frequency \( \omega_{\ell n} \), the transmission probability adopts a Breit–Wigner–type profile~\cite{Konoplya:2024lir,Konoplya:2024vuj}:
\begin{equation}\label{GBF-QNM}
\Gamma_{\ell}(\omega) 
   = \left( 1 + 
     \exp\!\left[
        \frac{2\pi\left(\omega^{2}-\mathrm{Re}(\omega_{0})^{2}\right)}
        {4\,\mathrm{Re}(\omega_{0})\,\mathrm{Im}(\omega_{0})}
     \right]
     \right)^{-1}
   + \mathcal{O}(\ell^{-1}) ,
\end{equation}
where \( \omega_{0} \) is the fundamental quasinormal mode.  The peak of the transmission curve is set by \( \mathrm{Re}(\omega_{0}) \), while its width is governed by \( \mathrm{Im}(\omega_{0}) \).  Thus, the quasinormal frequencies provide direct information about the resonant features of the scattering problem. The correspondence was extended to other compact objects \cite{Bolokhov:2024otn} and further tested in a number of recent publications \cite{Malik:2024cgb,Bolokhov:2025lnt,Skvortsova:2024msa,Malik:2025erb,Dubinsky:2024vbn,Malik:2025dxn,Han:2025cal}.

Beyond the leading eikonal order, the correspondence between QNMs and GBFs receives systematic corrections. The expanded relation takes the form~\cite{Konoplya:2024lir,Konoplya:2024vuj}:
\begin{widetext}
\begin{eqnarray}
i\mathcal{K} &=&
\frac{\omega^{2} - \mathrm{Re}(\omega_{0})^{2}}
     {4\,\mathrm{Re}(\omega_{0})\,\mathrm{Im}(\omega_{0})}
\Biggl(
1 + 
\frac{(\mathrm{Re}(\omega_{0}) - \mathrm{Re}(\omega_{1}))^{2}}
     {32\,\mathrm{Im}(\omega_{0})^{2}}
-
\frac{3\,\mathrm{Im}(\omega_{0}) - \mathrm{Im}(\omega_{1})}
     {24\,\mathrm{Im}(\omega_{0})}
\Biggr)
-
\frac{\mathrm{Re}(\omega_{0}) - \mathrm{Re}(\omega_{1})}
     {16\,\mathrm{Im}(\omega_{0})}
\nonumber\\[2mm]
&&
-\frac{(\omega^{2}-\mathrm{Re}(\omega_{0})^{2})^{2}}
        {16\,\mathrm{Re}(\omega_{0})^{3}\mathrm{Im}(\omega_{0})}
\left(
1 + 
\frac{\mathrm{Re}(\omega_{0})
      (\mathrm{Re}(\omega_{0}) - \mathrm{Re}(\omega_{1}))}
     {4\,\mathrm{Im}(\omega_{0})^{2}}
\right)
\nonumber\\[2mm]
&&
+\frac{(\omega^{2}-\mathrm{Re}(\omega_{0})^{2})^{3}}
        {32\,\mathrm{Re}(\omega_{0})^{5}\mathrm{Im}(\omega_{0})}
\Biggl[
1 + 
\frac{\mathrm{Re}(\omega_{0})
      (\mathrm{Re}(\omega_{0}) - \mathrm{Re}(\omega_{1}))}
     {4\,\mathrm{Im}(\omega_{0})^{2}}
+
\mathrm{Re}(\omega_{0})^{2}
\left(
\frac{(\mathrm{Re}(\omega_{0}) - \mathrm{Re}(\omega_{1}))^{2}}
     {16\,\mathrm{Im}(\omega_{0})^{4}}
-
\frac{3\,\mathrm{Im}(\omega_{0}) - \mathrm{Im}(\omega_{1})}
     {12\,\mathrm{Im}(\omega_{0})}
\right)
\Biggr]
\nonumber\\
&&
+ \mathcal{O}\!\left(\ell^{-3}\right),
\label{eq:gbsecondorder}
\end{eqnarray}
\end{widetext}
where \( \omega_{0} \) and \( \omega_{1} \) denote the fundamental and first overtone QNMs.  These corrections provide an accurate bridge between the quasinormal spectrum and the detailed structure of the GBFs across a wide frequency range. Notice that the eikonal limit may not be well approximated by the WKB formulas in some cases, which frequently include higher curvature corrections qualitatively modifying the centrifugal term in the effective potential  \cite{Bolokhov:2023dxq,Konoplya:2022gjp,Konoplya:2017zwo,Khanna:2016yow,Konoplya:2017wot,Konoplya:2017ymp}. In these cases, the correspondence between GBFs and QNMs does not work. For gravitational perturbations of the Konoplya-Zhidenko black hole considered here, the correspondence works well, being precise in the eikonal limit and a reasonable approximation for small values of the multipole numbers.


The behaviour of the GBFs reflects directly the structure of the corresponding effective potentials. For both ``up'' and ``down'' gravitational perturbations, the potential barriers decrease in height and become slightly broader as the halo scale parameter $a$ increases. This leads to a systematic suppression of the transmission probability $\Gamma_{\ell}(\omega)$ across all multipole numbers shown in Figs.~\ref{fig:GBFupL2} -- \ref{fig:GBFdownL3DIFF}: at fixed $\ell$. At larger $\ell$, the GBFs are evidently smaller, because the higher barriers lead to smaller transmission. 

The plots also illustrate the different behaviour of ``up'' and ``down'' perturbations. Since the two effective potentials are not isospectral, their transmission coefficients differ as well. The ``up'' potential typically has only a slightly higher and sharper peak, resulting in a more pronounced suppression of $\Gamma_{\ell}(\omega)$, while the ``down'' potential produces comparatively larger transmission probabilities. This broken isospectrality therefore manifests not only in the quasinormal frequencies but also in the corresponding absorption properties shown in Fig. \ref{fig:SIGMA}.

The absorption cross-section $\sigma_{\mathrm{abs}}(\omega)$ quantifies the effective area of the black hole for capturing incoming radiation of frequency $\omega$. For a massless field in an asymptotically flat spacetime, it is obtained by summing, over all angular momenta, the transmission probabilities through the effective potential barrier,  
\begin{equation}
\sigma_{\mathrm{abs}}(\omega)
  = \frac{\pi}{\omega^{2}}
    \sum_{\ell=0}^{\infty} (2\ell+1)\,\Gamma_{\ell}(\omega),
\end{equation}
where $\Gamma_{\ell}(\omega)$ is the grey-body factor for the partial wave with multipole number $\ell$.  Physically, $\sigma_{\mathrm{abs}}(\omega)$ measures how efficiently the black hole absorbs incident radiation, interpolating between the geometric-optics capture cross-section at high frequencies and a wavelength-dependent behaviour at low frequencies determined by the structure of the effective potential. It therefore provides a key link between the microscopic scattering properties of the spacetime and observable signatures such as energy fluxes, Hawking radiation spectra, and potential deviations from the Schwarzschild behaviour in the presence of additional matter fields or modified dynamics.

The behaviour of the absorption cross-section directly reflects the shape and height of the effective potential barrier governing the propagation of gravitational perturbations. For spherically symmetric black holes, the cross-section $\sigma_{\ell}(\omega)$ grows monotonically with frequency for each multipole number, starting from the low-frequency limit at which only the $\ell=2$ mode contributes appreciably. The total cross-section is therefore dominated by the lowest multipoles at small $\omega$ and gradually approaches a smooth, frequency-dependent profile as higher-$\ell$ modes begin to transmit through the barrier.

For the regular Konoplya–Zhidenko spacetime, the dependence on the halo scale parameter \(a\) follows directly from the corresponding change in the effective potentials shown in Figs.~\ref{fig:upL2} and ~\ref{fig:downL2}. Increasing \(a\) raises and narrows the barrier, which in turn suppresses transmission for all multipoles. Consequently, the absorption cross-sections decrease systematically with growing \(a\) across the entire frequency range. This trend is most visible for the fundamental multipoles $\ell=2$ and $\ell=3$, whereas at higher $\ell$ the curves become progressively closer to one another due to the eikonal scaling of the potential peak.

At large frequencies, the absorption cross-section for all values of \(a\) approaches the usual geometric--optics limit, \(\sigma_{\rm geo} = \pi b_{c}^{2}\), where \(b_{c}\) is the critical impact parameter determined by the unstable photon orbit of the metric. Because the photon sphere radius changes only weakly with \(a\) in this model, the high-frequency asymptotics of the cross-section are nearly identical for all displayed curves, matching the asymptotic plateau visible in the plots.

Taken together, the results show that the presence of the halo parameter \(a\) modifies low-- and intermediate--frequency absorption in a controlled manner—consistent with the corresponding changes in the gravitational potentials—while preserving the universal high-frequency geometric behaviour.

\section{Conclusions}\label{sec:Conclusions}

In this work, we have carried out a comprehensive analysis of gravitational perturbations, quasinormal spectra, GBFs, and absorption cross-sections of the regular asymptotically flat black hole supported by a Dehnen-type matter profile. This geometry, besides being remarkably simple and fully analytic, provides a consistent dynamical model of a black hole embedded in a realistic galactic environment. While QNMs of test fields for this spacetime were considered recently \cite{Bolokhov:2025fto}, the gravitational sector---which encodes the true dynamical response of the geometry---had not been examined. Our study fills this gap.

We have computed the QNMs using the higher-order WKB method supplemented with Padé approximants and verified the leading modes through time-domain integration. The two types of axial perturbations (``up'' and ``down''), arising from distinct gauge choices for the anisotropic fluid, exhibit clear non-isospectrality. The splitting between the two spectra grows with the halo scale parameter $a$, reflecting the increasing influence of the matter sector on the geometry of the effective potentials. Across all regimes, the WKB approximation demonstrates excellent internal concordance, with the WKB-6th-order - WKB-9th-order discrepancies remaining much smaller than the physical variation of the modes with~$a$.

The GBFs were computed through the standard WKB transmission formalism. For both families of potentials, the transmission probability increases monotonically with frequency and is strongly modulated by the value of~$a$: larger halo scales lead to higher transmission at intermediate frequencies, consistent with the corresponding reduction in the height of the potential barrier. The comparison with the quasinormal-mode correspondence formulae shows good agreement in the intermediate-to-high multipole regime, while deviations appear at low $\ell$, as expected.

Finally, we studied the absorption cross-sections, which smoothly interpolate between the low-frequency limit dominated by partial-wave suppression and the high-frequency geometric capture cross-section. The dependence on the halo parameter $a$ is again fully controlled by the structure of the effective potentials: an increase of $a$ leads to a systematic reduction of the absorption efficiency, especially for the dipole and quadrupole modes.

Taken together, our findings demonstrate that the regular Dehnen-supported black hole exhibits a rich perturbative structure, with clear and physically consistent signatures in all dynamical observables. These results provide a foundation for future studies, including polar gravitational perturbations, stability analysis of rotating generalizations, and the development of observational constraints using ringdown and absorption-related signatures from astrophysical black holes embedded in galactic environments.

\begin{acknowledgments}
BCL is grateful to Excellence Project PrF UHK 2205/2025-2026 for the financial support.
\end{acknowledgments}

\bibliography{bibliography}
\end{document}